\documentstyle[12pt]{article}
\begin{document}
\begin{flushright}
INR-0918/96 \\
February 1996\\
hep-ph/9603246  \\
\end{flushright}
\vspace*{1.0cm}
\begin{center}
{\bf \Large  QCD Corrections to Decays of Intermediate Mass \\
Higgs Boson} \footnote{To appear in {\it PNPI Research Report
 1994 -- 1995}, St.Petersburg} \\
\end{center}
\vspace*{1.0cm}
\begin{center}
{ \large Andrei L. Kataev$^{\ast}$ and Victor T. Kim$^{\dagger}$} \\
\vspace*{0.5cm}
{$^{\ast}$ Institute for Nuclear Research, 117312 Moscow, Russia}
\footnote{\it e-mail: kataev@ms2.inr.ac.ru; kataev@cernvm.cern.ch }

{$^{\dagger}$ St.Petersburg Nuclear Physics Institute, 1888350
Gatchina, Russia}
\footnote{\it e-mail: kim@pnpi.spb.ru }
\end{center}
\vspace*{2.0cm}

\begin{center}
{\bf \large Abstract}
\end{center}
\vspace*{0.5cm}
The brief discussion of the studies of the effects of the radiative
QCD corrections to the decay widths and branching ratios of the
Standard Model Higgs boson with the mass 
$2 m_b~GeV~\leq M_{H}\leq 2 M_W~GeV$ is
given. The numerical results are obtained with the help of the FORTRAN code
SEEHIGGS.

\newpage

 Among the most intriguing modern theoretical problems
are the investigations of the properties of still
non-discovered Higgs particles of the Standard Model (SM)
for ElectroWeak interactions. The special attention is paid
nowadays  to the searches of the SM Higgs boson.
Various aspects of "Higgs hunting" were discussed in detail from
the theoretical and the phenomenological
[1-3] points of view.

The current experimental lower bound on the SM
Higgs boson mass is about $M_H \geq 60$ $GeV$.
This lower bound came from the
 analysis of the LEP data taking
into account of the theoretical expression for $H\rightarrow q\overline{q}$
decay width with the QCD corrections of order
$O(\alpha_s)$
\cite{Bra80}.  The main decay channel of the SM Higgs boson in
the intermediate mass range  $2 m_b < M_H < 2M_W$ is the decay to the
$b\overline{b}$ final states with the coupling constant being
proportional to the $b$-quark mass. In this mass range the theoretical
uncertainties of $\Gamma(H \rightarrow b \overline{b})$
 are closely related to the theoretical uncertainties
of the branching ratios of some rare processes, e.g.,
  $H \rightarrow\gamma \gamma$
which is known as one of the most promising channels for the searches
of not too heavy Higgs bosons at Tevatron, LEP,
and soon at LEP200 and LHC.

 Here we present brief overview of current status of perturbative QCD
corrections to SM Higgs of intermediate mass.  We start from  analysis
of the perturbative QCD corrections for the Higgs decay into
$b$-quarks $\Gamma(H \rightarrow{b\overline{b}})$.
The width of this decay in Born level has form \begin{equation}
\Gamma(H \rightarrow{b\overline{b}}) \equiv \Gamma_{Hb\overline{b}} =
\Gamma_0^{(b)}\beta^3 \label{1} \end{equation}
where
$\Gamma
_{0}^{(b)} = \frac{3 \sqrt 2}{8\pi}G_F M_H m_b^2$,
$\beta = \sqrt{1-\frac{4m_b^2}{M_H^2}}$, $G_F$ -Fermi constant,
$M_H$ and $m_b$ are the pole masses of Higgs boson and $b$-quarks,
respectively.

The massless QCD corrections to $\Gamma(H^0\rightarrow q\overline{q})$ were
 considered  at the  next-to-next-to-leading order
 (NNLO) using the concept of the running $b$-quark mass
 $\overline{m}_b(M_H)$ \cite{Gor84},\cite{Gor90}
 In the process of the consideration of the  phenomenological
 consequences of the results of Ref.\cite{Gor90} it
 was noticed that  the variation of
 the running $b$-quark mass  from the $b$-quark mass-shell
 (namely from the $\overline{m}_b(m_b)$-value) to the
 $M_H$-scale (namely to the $\overline{m}_b(M_H)$-value)
 in the leading order (LO) of perturbation theory results in
 the negative corrections, which diminish the value of the
 corresponding Born expression for $\Gamma_{Hb\overline{b}}$
 by factor 2 \cite{Kle91}.

However, the running mass is not the unique way of defining mass
parameters. Indeed, in full analogy with the physical
mass of electron, in QCD one can
also define the pole (on-shell) quark mass.
The definition  of the pole quark mass is
commonly used for heavy quarks, namely for $c$- and $b$-quarks.  The
expression for the decay width $H \rightarrow q\overline{q}$
has been explicitly calculated in  terms of
the pole quark masses $m_c$ and $m_b$ in Ref.\cite{Bra80},
at the $O(\alpha_s)$ -level. The presented
in Ref.\cite{Gun90} numerical studies of these results  did not reveal
the effect of the $50\%$ reduction of the Born approximation.
Therefore, it is important to understand the origin of the observed in
Ref.\cite{Gun90} puzzle of the differences between various
parametrizations of the QCD results for $\Gamma_{H b\overline{b}}$ in
the experimentally interesting region of $M_H$ values.

Using the
2-loop relation between the running and the pole quark masses
\cite{Gra90} and the results of Ref.\cite{Gor90}
we have calculated the  expression for  $\Gamma_{Hb\overline{b}}$
 at the $\alpha_s^2$-level
in terms of the pole quark  mass in two different forms [9-11]. The
first one contains the $ln(M_H^2/m_b^2)$-contributions
explicitly, while in the second one they have been summed up through the
renormalization group (RG) technique. Note, that the second
parametrization is closely related to the one through the running
quark mass.

Our results demonstrate that for the non-RG-improved expression
for $\Gamma_{H b\overline{b}}$  calculated by us
$\alpha_s^2$-contribution  produce the
negative correction which is responsible for the elimination of
the numerical
difference between various parametrizations of
 the QCD results for
$\Gamma_{Hb\overline{b}}$ (see Fig.1,2, where
$R(H\rightarrow b\overline{b})=\Gamma_{Hb\overline{b}}/\Gamma_{0}^{(b)}$).

In [9-11] we analysed in detail the RG-improved
expression for $\Gamma_{Hb\overline{b}}$
taking into account  the effects, neglected in the course of
 estimates of Ref.\cite{Kle91}.   Among them are the corrections
 responsible for the relation between  the running
 mass $\overline{m}_b(m_b)$ and the pole mass $m_b$ at the
 1-loop and 2-loop levels and the order $O(m_b^2/M_H^2)$-corrections to
 $\Gamma_{Hb\overline{b}}$. The importance of taking
 into account of the order $O(m_b^2/M_H^2)$-corrections for modeling
 the threshold effects was demonstrated in Refs.[9-11].
Following the considerations of Ref.\cite{KatPL} we also calculated
analytically the mixed QED- and QCD- corrections of the
order $O(\alpha\alpha_s)$ to
$\Gamma_{Hb\overline{b}}$ \cite{Kat92}, which turned out to be very small.

Taking into account the discussed above
 QCD uncertainties and the current uncertainties of the value  of the
 parameter $\Lambda_{\overline{MS}}^{(5)}$  we obtain the following
 theoretical estimate for $\Gamma_{Hb\overline{b}}$ in the intermediate
 region of $M_H$ values $60\ GeV <M_H< 160  \ GeV$ (see Fig.1,2)
\begin{equation}
\Gamma_{Hb\overline{b}}=\bigg(0.55\div
0.45\bigg) \Gamma_{0}^{(b)} \label{h}
\end{equation}
which confirms the rough estimation presented in \cite{Kle91}.

Therefore we confirmed also observation \cite{Kle91} that the
discussed QCD contributions to
$\Gamma_{Hb\overline{b}}$ are increasing the values of the branching
ratios of the decays $H^0\rightarrow l^-l^+$ and
 $H^0\rightarrow\gamma\gamma$ by the factor of over 2 (see Fig.3)
for the intermediate
mass SM Higgs because the QCD corrections for them are
small (few percents) \cite{Djo91}. Note, that all figures,
presented here were plotted using the values of the pole quark
masses $m_b=4.6\ GeV$ and $m_c=1.4\ GeV$ and
$\Lambda_{\overline{MS}}^{(5)}=250\ MeV$.
The detailed consideration of the sensitivity of the results obtained to
the variation of $\Lambda_{\overline{MS}}^{(5)}$ value up to over
$150\ MeV$ is under study.

Note, that there is still room for curious speculation \cite{Kat94}
on possible consequences of the observed property of the asymptotic
explosion of the considered by us NNLO parametrization of
$\Gamma_{Hb\overline{b}}$.
 We think it might indicate a possible existence
of some resonance at the mass scale of $M_{ch} =
60 \div 130$ GeV (for $\Lambda_{\overline{MS}}^{(5)}=250 \div 150$ MeV).

The results of our analysis [9-11] were taken into
account in the series of the subsequent calculations
[14-16].  In Ref.\cite{Sur94} the analyzed by us
uncertainties due to the inclusion of the effects of the order
$O(\alpha_sm_b^2/M_H^2)$-corrections were fixed by the explicit
analytical calculations, while in Ref.\cite{Che95} the
mentioned by us contribution to $\Gamma_{Hb\overline{b}}$ of the
triangle-type diagram with the virtual top-quark was evaluated. The
large top-quark mass expansion for $H\rightarrow b\overline{b}$
decay rate in the order $\alpha_s^2$ and for $H\rightarrow gluons$
decay rate in order $\alpha_s^3$ were calculated in
 Ref.\cite{Lar95}.

The outcomes of our studies of
Refs.[9-11]
 are forming the basis for the FORTRAN
code SEEHIGGS, which includes the results of the most recent
calculations of Refs.[14-16]
and is under development.

This work on this brief report was done within the project
N96-02-18897 of the Russian Foundation for Fundamental
Research.


\begin{thebibliography}{10}
\bibitem{Vel93} M. Veltman, {\it The Higgs System},
In: {\it Perspectives in Higgs Physics},
Ed. G.L.Kane, World Scientific, Singapore (1993), pp.1-45;
Instituut-Lorenz report No.0601, Leiden (1989)
\bibitem{Gun90} J.F.Gunion, H.E.Haber, G.L.Kane and S.Dawson,
 \newblock {\it The Higgs Hunter's Guide}, Addison-Wesley
 Publishing Co. (1990)
\bibitem{Hab96} H.E.Haber, Invited talk
 at the Int. Europhysics Conf.  on High Energy Physics, 1995,
 Brussels, CERN-TH/96-07 (1996); A.Djouadi, M.Spira, and
P.M.Zerwas, preprint DESY 95-210 (1995);
 B.A.Kniehl, {\it Int. J. Mod. Phys.} {\bf A10} (1994) 443;

\bibitem{Bra80} E.Braaten
and J.P.Leveille, \newblock{\it Phys.  Rev.} {\bf D22} (1980) 715;
M.Drees and K.Hikasa, \newblock{\it Phys. Rev.} {\bf D41}
 (1990) 1547; \newblock{\it Phys. Lett.} {\bf B240} (1990) 455; ibid.
 {\bf B262} (1991) 497 (E)
\bibitem{Gor84} S.G.Gorishny,
A.L.Kataev and S.A.Larin, \newblock{\it Sov. J. Nucl. Phys.} {\bf 40}
 (1984) 329
\bibitem{Gor90} S.G.Gorishny, A.L.Kataev, S.A.Larin
 and L.R.Surguladze, \newblock{\it Mod. Phys. Lett.} {\bf A5} (1990)
 2703; \newblock{\it Phys. Rev.} {\bf D43} (1991) 1633
\bibitem{Kle91}
 R.Kleiss, Z.Kunszt and W.J.Stirling, \newblock{\it Phys. Lett.} {\bf
 B253} (1991) 269
 \bibitem{Gra90}
 N.Gray, D.J.Broadhurst, W.Grafe and K.Schilcher,
 \newblock{\it Z. Phys.} {\bf C48} (1990) 673
 \bibitem{Kat92}
A.L.Kataev and V.T.Kim, ENSLAPP-A-407/92, Annecy (1992)
\bibitem{Kat93}
A.L.Kataev and V.T.Kim,
In:  {\it Proc. III Int. Workshop on Software Engineering,
Artificial Intelligence and Expert Systems in High Energy
and Nuclear Physics}, October 1993, Oberammergau, Germany,
Eds. K.-H.Becks and D.Perret-Gallix,
World Scientific, Singapore (1994), pp.623-637
\bibitem{Kat94}
A.L.Kataev and V.T.Kim,
{\it Mod. Phys. Lett.} {\bf A9} (1994) 1309
\bibitem{KatPL} A.L.Kataev, {\it Phys. Lett.} {\bf B287} (1992) 209
\bibitem{Djo91}
A.Djouadi, M.Spira and P.M.Zerwas,
{\it Phys. Lett.} {\bf B264} (1991) 440
\bibitem{Sur94}
L.R.Surguladze, {\it Phys. Lett.} {\bf B341} (1994) 61
\bibitem{Che95}
K.G.Chetyrkin and A.Kwiatkowski,
LBL-37269, Berkeley (1995)
\bibitem{Lar95}
S.A.Larin, T. van Ritbergen and J.A.M.Vermaseren,
{\it Phys. Lett.} {\bf B362} (1995) 134


\end{thebibliography}
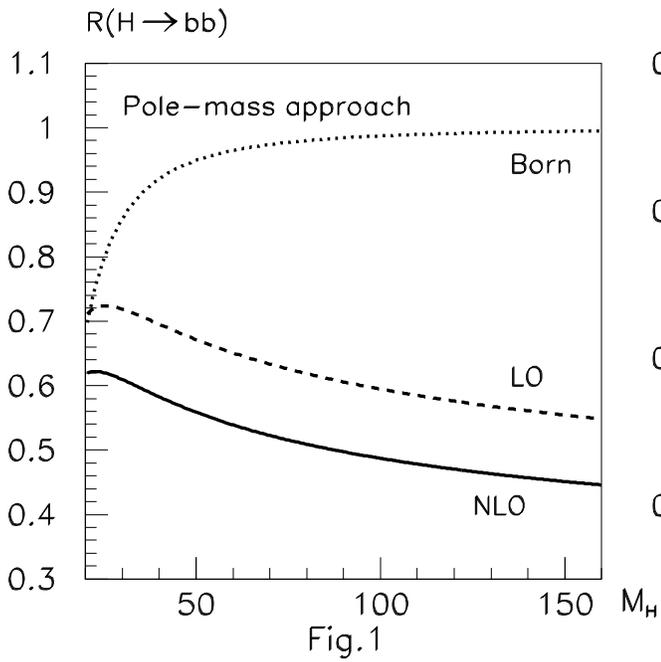
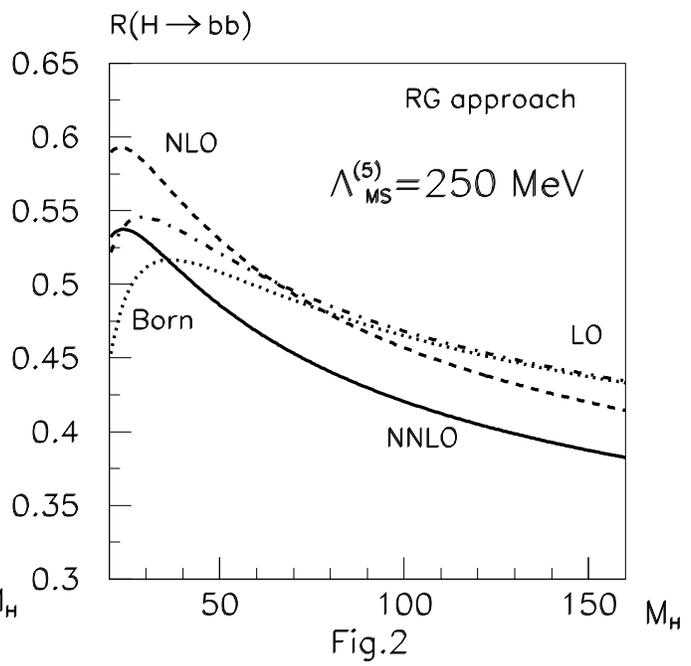
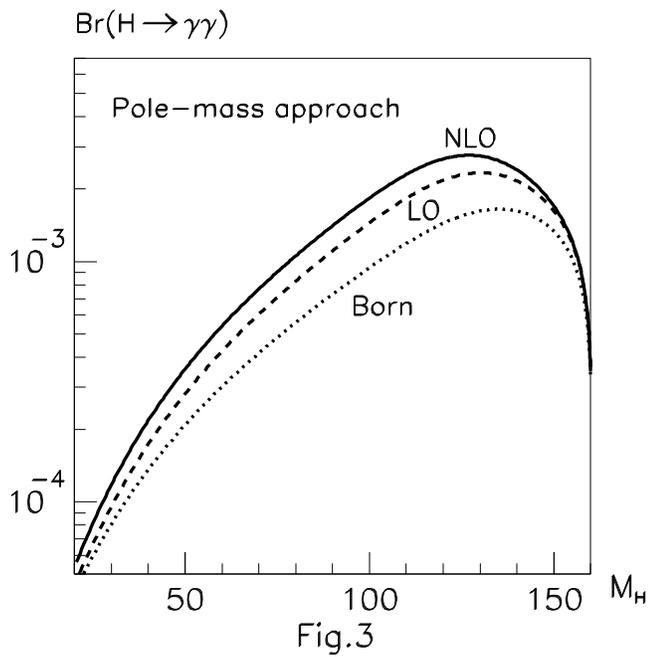
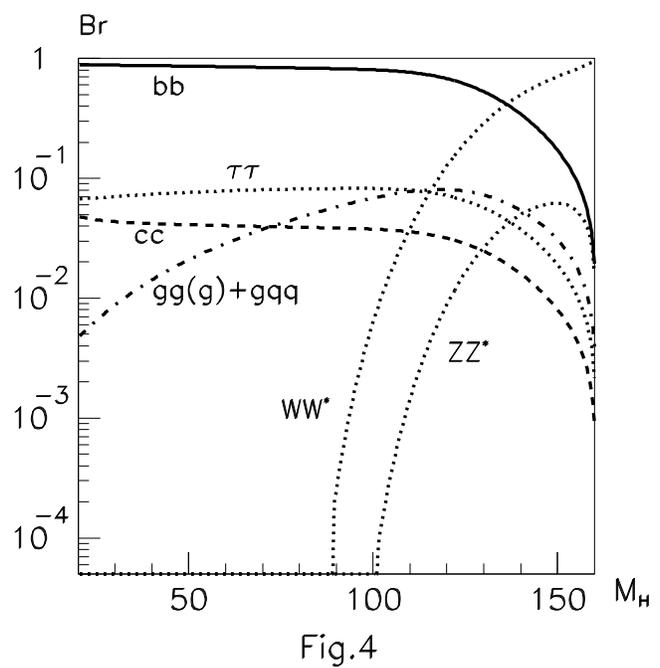
\end{document}